\documentclass[a4paper]{jpconf}
\usepackage{graphicx}
\begin{document}
\def\br{\begin{eqnarray}}
\def\er{\end{eqnarray}}
\def\be{\begin{equation}}
\def\ee{\end{equation}}
\def\({\left(}
\def\){\right)}
\def\d{\delta}
\def\h{ {1\over 2}  }
\def\m{\mu}
\def\o{\over}

\title{Quantum Confinement in Hydrogen Bond of DNA and RNA}
\author{C S dos Santos, E Drigo Filho}
\address{Instituto de Bioci\^encias, Letras e Ci\^encias
Exatas, IBILCE-UNESP\\ Rua Cristov\~ao Colombo 2265, 15054-000 S\~ao Jos\'e do Rio Preto ,SP, Brazil}
\ead{ carlos.sisantos@catolica.edu.br, drigo@ibilce.unesp.br}
\author{R M Ricotta}
\address{Faculdade de Tecnologia de S\~ao Paulo, FATEC/SP-CEETPS-UNESP
\\ Pra\c ca  Fernando Prestes 30,  01124-060 S\~ao Paulo, SP, Brazil}
\ead{regina@fatecsp.br}
\begin{abstract} 
The hydrogen bond is a fundamental ingredient to stabilize the DNA and RNA macromolecules. The main contribution of this work is to describe quantitatively this interaction as a consequence of the quantum confinement of the hydrogen. The results for the free and confined systems are compared with experimental data. The formalism to compute the energy gap of the vibration motion used to identify the spectrum lines is the Variational Method allied to Supersymmetric Quantum Mechanics.
 
\end{abstract}

\section{Introduction}
One of the most important interactions to stabilize the DNA and RNA macromolecules is the hydrogen bond between pairs of nitrogenous bases. It  promotes a link among the complementary pairs of bases, maintaining the double helix structure. In the DNA the pairs of bases are guanine-cytosine and adenine-thymine; in the RNA the thymine is substituted by uracil, \cite{Yang}. In a recent work, the hydrogen bonds present in proteins and in smaller molecules were studied by considering the hydrogen  quantum confinement in the NH and OH molecular groups, \cite{Santos}. In the present work, we extended the results for the nucleic acids, which present the NH molecular group. 

The hydrogen bond is present in various systems,  \cite{Jeffrey}, and it is classified as a donor-acceptor interaction. From the experimental point of view, it is possible to identify if a particular molecular group is forming an hydrogen bond due to the change in the vibrational spectrum. This change can be observed by different experimental techniques, as for instance the infrared spectrum analysis. 

The main point of this work is to show that the hydrogen bond can be treated as the quantum confinement of the hydrogen, which would change the transitions related to the NH group present in nucleic acids. As usual, the Morse potential, \cite{Morse}, is used to simulate the vibration of the NH molecular group,  \cite{Leviel}-\cite{Blaise}. The parameters used are the ones found in literature for these groups without the hydrogen interaction, i.e., without confinement. The geometrical restrictions due to the hydrogen bonds are obtained from standard experimental results for this kind of systems, \cite{Jeffrey}. The calculation of the energy levels is made through the variational method with test functions obtained from Supersymmetric Quantum Mechanics formalism, SQM. This procedure has alrealdy been used in different quantum problems, with and without  confinement,  \cite{Drigo2}-\cite{Varshni}, including specific studies involving the Morse potential, \cite{Ley-Koo}, which corroborates its  applicability  in the present case.

In section 2 the adopted model for the hydrogen bonds is presented. We then present a review of the SQM formalism in section 3. In Section 4  the specific mathematical formalism concerning the Morse potential is developed, including the identification of the wave functions to be used in the Variational Method. The numerical results, with the appropriate parameters of the NH group are presented as well as the comparison with the experimental data are in section 5. In section 6  the conclusions are discussed.

\section{Model}
Figure 1 shows a schematic model of a hydrogen bonding between the donor atom X and the acceptor atom Y, here denoted by  X-H$\cdots$Y.  The hydrogen bond is represented by H$\cdots$Y and covalent bond is represented by X-H. The X and Y atoms are considered as hard spheres and the hydrogen is treated as a material point, since the electron cloud is shared with the donor and the acceptor atoms. The origin of the coordinate system is located at the center of the atom X. Any distance used is computed from this point and it is considered center-to-center. The covalent radius is adopted by the condition of impenetrability of the donor atom X; the employment of the van der Waals radius is justified because it is a measure of the volume excluded by the acceptor atom Y, i.e., the hydrogen cannot penetrate this region. 

Thus, the hydrogen nuclei may oscillate within a region determined between a minimum distance ($x_{min}$), given by the covalent radius ($x_c$) of atom X, and a maximum distance ($x_{max}$), given by the length of the hydrogen bond ($l_{x \cdots y}$) subtracted by the van der Waals radius ($x_{vw}$) of the atom Y, i.e.: 
\be
\label{xmax}
x_{max} = l_{X \cdots Y} - x_{vw}
\ee
\begin{figure}
\centering
\includegraphics[width=0.6\textwidth]{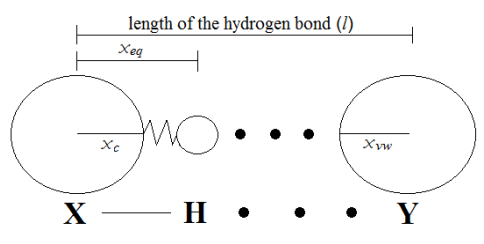}
\caption{\label{label} Schematic one-dimensional model, where $x_c$ is the covalent radius of the atom donor, $x_{vw}$ is the van der Waals radius of the atom and acceptor and $x_{eq}$ is the equilibrium distance of covalent bond. X is the donor atom and Y represents the atom acceptor. }
\end{figure}

The vibrational energy of this  quantum system is suitably described by the confined one-dimensional Morse potential,   widely known to well describe oscillations of diatomic molecules, \cite{Morse}, \cite{Drigo2}-\cite{Silva}, \cite{Ley-Koo}. The energy absorbed/emitted is obtained from the difference between the energies of the ground state and the first excited state. Since the confined Morse potential is non-exactly solvable, the energies of those states evaluated by using the variational method associated to the formalism of SQM.

\section{Supersymmetric Quantum Mechanics, SQM}
SQM is a well known algebraic method commonly used to exploit different aspects of non-relativistic quantum mechanical systems. It is particularly efficient to solve exactly all the quantum potential problems by means of the Hamiltonian factorization, thus providing the entire spectrum of wave functions and respective energies and a hierarchy of Hamiltonians, all related by the supersymmetric algebra, \cite{Sukumar}. 

Using the superalgebra,  a given Hamiltonian $H_1$, which has the energy $E_0^{(1)} $ and wave function $\Psi_0^{(1)}(x)$ for the lowest state, can be factorized in terms of bosonic operators $A_1^{\pm} $, such that the   Schr$\ddot {o}$dinger equation can be rewritten (in $\hbar = c = 1$ units) as

\be
\label{H1}
H_1 =  -{d^2 \o d x^2} + V_1(x) =  A_1^+A_1^-  + E_0^{(1)} 
\ee
where
\be
A_1^- \Psi_0^{(1)}(x)=0.
\ee
The following realization of the bosonic operators in terms of the superpotential $w_1(x)$ 
\be 
A_1^{\pm} =  \left(\mp {d \o dx} + w_1(x) \right) 
\ee
results in the Riccati equation
\be
\label{Riccati}
w_1^2 - {d \o dx}w_1=  V_1(x) - E_0^{(1)}   
\ee
and defines the ground state wave function in terms of the superpotential
\be
\label{eigenfunction}
\Psi_0^{(1)} (x) = N exp( -\int_0^x w_1(\bar x) d\bar x).
\ee

It can be shown that a $n$-members hierarchy of Hamiltonians can be generated, all related by the superalgebra, \cite{Sukumar},
\be
H_n = A_n^+A_n^- + E_0^{(n)} \;\;,
\ee
where 
\be 
A_n^{\pm} =  \left(\mp {d \o dx} + w_n(x) \right) 
\ee
with simple relations connecting the eigenvalues and eigenfunctions of the $n$-members, 
\be
\label{Psis}
\Psi_n^{(1)} = A_1^+A_2^+...\psi_0^{(n+1)}\;,\;\;\;\;E_n^{(1)} = E_0^{(n+1)}.
\ee
In equation (\ref{Psis}), the upper index between parentheses refers to the hierarchy Hamiltonian member and the lower index to the level within the hierarchy. 

At this point, it should be remarked that if the potential is non-exactly solvable, such as in a confined system, an approximation technique is needed and the variational method has already appeared as fully appropriate, \cite{Drigo2}.

\section{Mathematical Formalism}

The one-dimensional Morse potential, \cite{Morse}, defined in terms of the parameters $D_e $, the dissociation energy of molecule; $\beta$,  the parameter related to the width of potential well and $x_{eq}$, the inter-nuclear equilibrium distance, is given by
\be
V_M = D_e (e^{-2\beta(x - x_{eq})} - 2e^{-\beta(x - x_{eq})}).
\ee
By treating the molecule as an oscillator of frequency $\nu$, we can Taylor expand the Morse potential around the equilibrium position to find, in second order, the $\beta$ parameter given by 
\be
\label{beta}
\beta= ({2\pi^2 \mu\over D_e})^\h\nu
\ee
and fix it through experimental data. With the  change of variables, $y = \beta x$, the dimensionless Schr$\ddot {o}$dinger equation for a molecule subject to this potential is given by 
\be
\label{SE}
-{d^2 \Psi(y) \over dy^2} + \Lambda^2(e^{-2(y- y_{eq})} - 2e^{-(y - y_{eq})})\Psi(y) = \epsilon \Psi(y)
\ee
with
\be
\label{lambda}
 \Lambda^2= {2 \mu D_e\over \beta^2\hbar^2}
\ee
and 
\be
 \epsilon= {2 \mu E\over \beta^2\hbar^2}
\ee
where $\mu$ is the reduced mass of the system. The Hamiltonian operator in equation (\ref{SE}) is given by
\be
\label{Morse-Hamiltonian}
\hat H_M = -{d^2  \over dy^2} + \Lambda^2(e^{-2(y- y_{eq})} - 2e^{-(y - y_{eq})})
\ee
Since  the Schr$\ddot {o}$dinger equation  is exactly solvable, we identify $\hat H_M$ with $H_1$ of equation (\ref{H1}) and proceed the process of factorization, calculating the whole hierarchy. The
result is, \cite{Drigo2}
\br
V_{n+1}(y) = \Lambda^2( e^{-2(y - y_{eq})} - 2e^{-(y - y_{eq})}) + 2n \Lambda e^{-(y - y_{eq})}
\nonumber
\er
\br
w_{n+1}(y) = - \Lambda e^{-(y - y_{eq})} + (\lambda - {2n + 1\o 2})  \nonumber
\er
\br
\label{non-confined energy}
\epsilon_0^{(n+1)} = -(\Lambda - {2n + 1\o 2})^2. 
\er
\\
In addition we can calculate exactly the wave functions. The ground state wave function is evaluated by using equation (\ref{eigenfunction}) and is given by 
\be
\label{Psi1}
\Psi_0^{(1)}(y) \propto exp\left( {-\Lambda e^{-(y - y_{eq})}\;  -y(\Lambda -\h)} \right)
\ee
This wave function is then used to evaluate, through the SQM algebra,  the  wave function for the excited states by using equation (9); the first excited state is  given by
\be
\label{Psi2}
\Psi_1^{(1)}(y) \propto -2 \left( {\Lambda e^{-(y - y_{eq})}-\Lambda +1} \right) exp\left( {-\Lambda e^{-(y - y_{eq})}\;  -y(\Lambda - {3\o 2})} \right)
\ee
The wave functions given in (\ref{Psi1}) and (\ref{Psi2}) are the ground state and the first excited state wave functions of the exact system, non-confined or with no hydrogen bond. We refer to them as $ {\Psi_n}_{(non-confined)}$, $n=0,1$. As our aim is the lowest state and the first excited state of the confined system, with the hydrogen bond, a potential barrier is introduced, i.e., $V_1(y_{max})=V_1(y_{min}) = \infty$, meaning that these wave functions must be zero outside the barrier. Consequently, the probability of finding the hydrogen outside the confinement region must be zero. 

As the system is no longer exact, it  is necessary to use an approximative method, which we choose to be the variational method associated to SQM. From the SQM point of view, this is equivalent to introduce two additional terms in the superpotential $w_1$, associated to the superalgebra, \cite{Drigo1},  \cite{Silva}.  Adopting this procedure, the wave functions to be used in the variational method are given by
 \be
 \label{Psi-confined}
{\Psi_n^{(1)}}_{(confined)} \propto {\Psi_n^{(1)}}_{(non-confined)} \; .(y_{max} - y)(y-y_{min})\;\;\;n=0,1\;\;,
\ee
with the corresponding energies  obtained by solving
\be
\label{energylevels-confined}
E_n = {\int_{y_{min}}^{y_{max}}{\Psi_n^* \hat H_M \Psi_n dy}\over  {\int_{y_{min}}^{y_{max}}{\Psi_n^*\Psi_n dy}}}
\ee
where $\hat H_M$ is given by equation (\ref{Morse-Hamiltonian}) and $\Psi_n$  are the trial wave functions given by equation (\ref{Psi-confined}), which depend on (\ref{Psi1}) and  (\ref{Psi2}). The integration limits are the covalent radius of the atom for which the hydrogen is chemically bound ($x_{min}$) and the van der Waals radius complementary of hydrogen bond ($x_{max}$). The results of energy are presented in $cm^{-1}$ units (inverse wavelength), the same as the ones from spectroscopic results.

\section{Results}

\begin{figure}
\begin{center}
\includegraphics[width=0.4\textwidth]{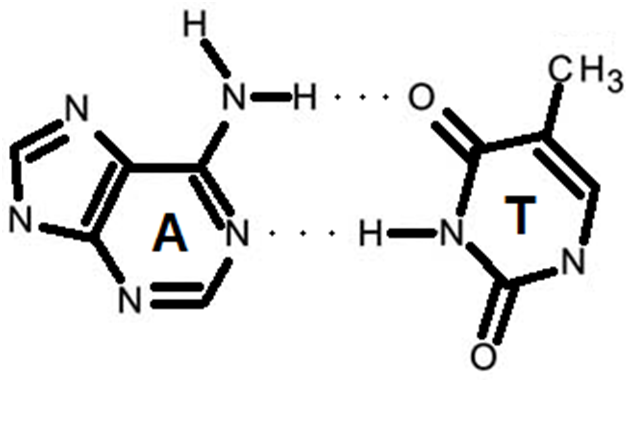}
\caption{\label{label} Schematic model hydrogen bonds present in the AT pair in DNA.}
\end{center}
\end{figure}

\begin{figure}
\begin{center}
\includegraphics[width=0.4\textwidth]{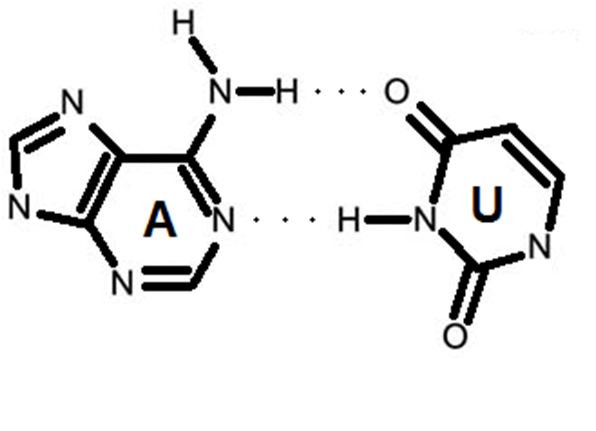}
\caption{\label{label} Schematic model hydrogen bonds present in the AU pair in RNA.}
\end{center}
\end{figure}
The hydrogen bonds studied involve the NH group. The necessary parameters, presented in Tables 1 and 2 below, were obtained experimentally, as indicated.
\vskip .3cm
\noindent {\bf Table 1.} Parameters used for the calculations of energy to the NH group, \cite{Janoschek}.

\vskip .3cm
\begin{center}
\noindent \begin{tabular}{lcc} \hline
\multicolumn{1}{l} { } & 
\multicolumn{1}{c} {$NH$} &\\ \hline \\
Equilibrium distance ($x_{eq}$)& 1.0362 \AA\\  
Harmonic vibrational wavenumber ($\omega$)& $3282 cm^{-1}$\\ 
Energy dissociation ($D_e$) & 3.47 eV\\  
Reduced mass ($\mu$)&$1.5614.10^{-27} kg$  \\ \hline 
\end{tabular}\\
\end{center}

\vskip .3cm

\noindent {\bf Table 2.} Covalent radius and van der Waals radius used for the nitrogen and oxygen atoms \cite{Cordero}, \cite{Bondi}.

\vskip .3cm
\begin{center}
\noindent \begin{tabular}{lcc} \hline
\multicolumn{1}{l} { } & 
\multicolumn{1}{c} {$N$} &
\multicolumn{1}{c} {$O$} \\ \hline 
Covalent radius ($x_c$)& 0.71 \AA& 0.66 \AA\\  
van der Waals radius ($x_{vw}$) & 1.55\AA &1.52 \AA\\  \hline

\end{tabular}\\
\end{center}
\vskip .3cm
\subsection{Results for NH group}
\vskip .3cm
\subsubsection{Non-confined case} \hfill
\vskip .3cm
Using the values given in Table 1 the value of the Morse potential $\beta$ parameter, determined by the relation (\ref{beta}) for NH group is $\beta = 2.3168 .10^{10} m^{-1}$. The value of the constant $\Lambda$ for NH group, given by equation (\ref{lambda}), is $\Lambda = 17.0534$.

In the situation free of any confinement the calculation of the transition from the first excited state to the ground state is made according to equation (\ref{non-confined energy}). In this case, in spectroscopic units, this value corresponds to $E_1-E_0 = 3089 cm^{-1}$ and should be found in spectroscopy. In studies by infrared spectroscopy with polypeptides it is observed an absorption band in $3090 cm^{-1}$ and it has been described as a vibration frequency of the NH  molecular group, \cite{Badger}. This value is found very close to the calculated and can be identified, according to the model proposed, as vibration of NH without the formation of hydrogen bonding.

\subsubsection{Confined case} \hfill
\vskip .3cm
When the NH group form molecular hydrogen bonding the vibration frequency is shifted. An interesting system where you can observe this effect are in the nucleic acids, DNA and RNA, because between the base pairs there are hydrogen bonds that allow the maintenance of the structure of the macromolecule. Here we analyze the vibrations of the NH group in hydrogen bonding between base pairs adenine-thymine (AT) and adenine-uracil (AU). In these base pairs two hydrogen bonds are formed, one with oxygen and the other with nitrogen. The schematic models are shown in Figures 2 and 3.

The lengths of the hydrogen bonds of AT and AU present in the DNA and RNA molecules, respectively, are well known in the literature and are obtained by X-ray diffraction, \cite{Saenger}. The distances are given between the donor atom and acceptor atom, because by X-ray diffraction it is not possible to determine the hydrogen positions. The lengths of the hydrogen bonds are shown in Table 3 below. 
\vskip 0.3cm
\noindent {\bf Table 3.} Length of hydrogen bonds in nucleic acids, \cite{Saenger}.
\begin{center}
\noindent \begin{tabular}{lccccccccc} \hline
\multicolumn{1}{c} { } & 
\multicolumn{1}{c} { } &  
\multicolumn{1}{c} {AT} &
\multicolumn{1}{c} { } & 
\multicolumn{1}{c} { } & 
\multicolumn{1}{c} {AU}  \\ \hline 
(N...N 	&&2.927 \AA&&&	2.820\AA \\  \hline
(N...O) 	&&2.861 \AA&&&	2.950\AA\\  \hline
\end{tabular}\\
\end{center}
\vskip 0.3cm

These lengths are used in equation (\ref{xmax}) to determine the hydrogen maximum limit of oscillation.

Solving equation (\ref{energylevels-confined}) through the variational method, we found the transition energies for each bond; they are summarized in Table 4. These values are compared with results given in literature obtained experimentally.
\vskip .3cm

\noindent {\bf Table 4.} Energy levels of the NH group in the hydrogen bonds between the base pairs AT and AU. $E_0$ is the ground state energy, $E_1$ is the energy of the first excited state and  $\Delta E_{1\rightarrow 0}= E_1- E_0$, in $(cm^{-1})$ units.
\begin{center}
\noindent \begin{tabular}{lcccccccc} \hline
\multicolumn{1}{c} { } & 
\multicolumn{1}{c} { } & 
\multicolumn{1}{c} { } & 
\multicolumn{1}{c} { } & 
\multicolumn{1}{c} {$E_0$} &
\multicolumn{1}{c} { } & 
\multicolumn{1}{c} {$E_1$} & 
\multicolumn{1}{c} { } & 
\multicolumn{1}{c} {$\Delta E_{1\rightarrow 0}$} \\ \hline
AT&& (N$\cdots$N) && -26330 & & -23122  &&  3218\\  
 && (N$\cdots$O) && -26314 & &	-23039	&& 3275\\  \hline
 
 AU&&	(N$\cdots$N)	&& -26225 & &	-22794 &&	3431\\
	&& (N$\cdots$O)&&	-26340	& & -23175 &&	3165\\ \hline
 
\end{tabular}\\
\end{center}
\vskip .3cm

Femtosecond vibrational spectroscopy with DNA oligomers realized with an alternating sequence of 23 adenine and thymine base pairs,  \cite{Szyc}, shows that there are three absortion peaks in the vibrational spectrum interesting to remark.  

One is between $3000$ and $3100 cm^{-1}$ and two around $3250 cm^{-1}$, one being located between $3200$ and $3250 cm^{-1}$, related to the N...N  hydrogen bond in the AT pair, and the other between $3250$ e $3300 cm^{-1}$, related to the N...O  hydrogen bond in the AT pair. These last two peaks can be identified, according to our model, as the vibration of the hydrogen bond in NH, with a deviation between  $0.56\%$-$0.98\%$ and $0.76\%$-$0.77\%$ respectively to the two ranges.  The first peak, located around $3000$ and $3100 cm^{-1}$, was found to be in fact at $3075cm^{-1}$,  and can be associated, according to our model ($3089cm^{-1}$), with NH group without hydrogen bond, with a deviation of $0.46\%$. 

Thus, these experimental results of the AT of DNA and AU of RNA oligomers show perfect agreement to the results obtained through our method.

Other infrared spectroscopy studies with RNA oligomers, \cite{Woutersen}, show the vibration spectrum when the NH participates in hydrogen bonds with some explicit numerical values. One of the peaks of the spectrum is located at $3185 cm^{-1}$, very close to $3165 cm^{-1}$ relative to the binding (N$\cdots$O) calculated by our method, with a deviation of  only $0.63\%$. A peak between  $3050$ e $3100 cm^{-1}$ can also be observed, which can de associated to the non-confined situation, with no hydrogen bond  of NH group of our model, ($3089cm^{-1}$).

\section{Conclusion}

In this work the quantum confinement of hydrogen present when hydrogen bonds are stablished was studied through the variational method associated to SQM.   The vibration spectrum of NH group was evaluated in the situations of non-confinement, i.e., with no hydrogen bonds and when there is confinement, i.e., when the hydrogen bonds are formed. We focused our attention to the case of DNA and RNA oligomers, which present the NH group. As a result we conclude that the presence of hydrogen bonds promote  significant changes in the vibration spectrum of NH group.  The comparison with experimental data reported in the literature showed perfect agreement, with  percentage errors smaller than $1\%$, which reenforces the applicability of the proposal. 

We conclude that the variational method associated to SQM was successfully applied to the calculation of the vibrational spectrum of NH molecular group with the principle of quantum confinement of hydrogen bonds. The treatment proposed here can be extended to the study of other molecular groups, as FH and CH, involved in hydrogen bonds.

\end{document}